\documentclass[]{spie}

\usepackage[]{graphicx}
\usepackage{caption}
\usepackage{subcaption}
\usepackage{xfrac}

\def  \be {\begin{equation}}
\def  \ee {\end{equation}}

\title{ A Novel Reflectometer for Relative Reflectance Measurements of CCDs }

\author{Murdock  Hart\supit{a}, Robert H. Barkhouser\supit{a}, James E. Gunn\supit{b}, and Stephen A. Smee\supit{a}
\skiplinehalf
\supit{a}Department of Physics and Astronomy, Johns Hopkins University, Baltimore, MD, 21218 (USA); \\
\supit{b}Department of Astrophysical Sciences, Princeton University, Princeton, NJ, 08544 (USA)
}

\begin{document} 
\maketitle


\begin{abstract}

The high quantum efficiencies (QE) of backside illuminated charge coupled devices (CCD) has ushered in the age of the large scale astronomical survey. The QE of these devices can be greater than 90\%, and is dependent upon the operating temperature, device thickness, backside charging mechanisms, and anti-reflection (AR) coatings. But at optical wavelengths the QE is well approximated as one minus the reflectance, thus the measurement of the backside reflectivity of these devices provides a second independent measure of their QE. 

We have designed and constructed a novel instrument to measure the relative specular reflectance of CCD detectors, with a significant portion of this device being constructed using a 3D fused deposition model (FDM) printer. This device implements both a monitor and measurement photodiode to simultaneously collect incident and reflected measurements reducing errors introduced by the relative reflectance calibration process. While most relative reflectometers are highly dependent upon a precisely repeatable target distance for accurate measurements, we have implemented a method of measurement which minimizes these errors. 

Using the reflectometer we have measured the reflectance of two types of Hamamatsu CCD detectors. The first device is a Hamamatsu 2k x 4k backside illuminated high resistivity p-type silicon detector which has been optimized to operate in the blue from 380 nm - 650 nm. The second detector being a 2k x 4k backside illuminated high resistivity p-type silicon detector optimized for use in the red from 640 nm - 960 nm. We have not only been able to measure the reflectance of these devices as a function of wavelength we have also sampled the reflectance as a function of position on the device, and found a reflection gradient across these devices.

\end{abstract}



\section{Reflectometer Device}
\label{device}

The reflectometer was designed to measure the specular reflectance relative to a silicon calibration standard at wavelengths from 380 nm to 1100 nm. To perform a measurement a collimated beam of light 8 mm in diameter is incident upon a target surface at an angle $15^{\circ}$ from normal. The effects upon the reflectance due to polarization can be neglected as the collimated beam is not polarized and these effects are also negligible at a $15^{\circ}$ angle from normal incidence, the reflectance is taken to be the average of the s and p polarizations. 

\begin{figure}[h]
	\hspace{1cm}
	\includegraphics[width=0.8\textwidth]{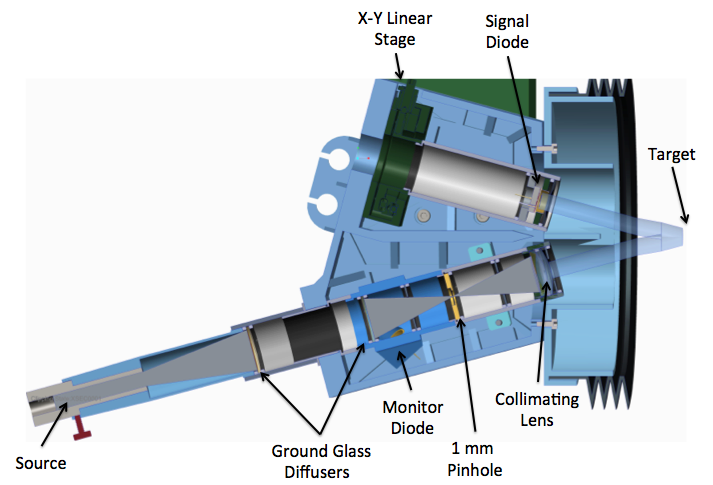}
	\caption{ Reflectometer cross section view. A slit to round fiber couples the reflectometer to a monochromator source. The fiber illuminates a pair of ground glass diffusers to create an even illumination pattern. The second ground glass diffuser is imaged by a pinhole, and the pinhole is imaged by an aspheric lens. The aspheric lens creates a collimated beam which is incident upon the surface to be measured $15^{\circ}$ from normal. Finally the collimated beam is reflected back to a signal photodiode. A monitor diode is placed to measure the output of the second ground glass diffuser. }
	\label{reflectometer_xsec}
\end{figure}

The main structure of the reflectometer was 3D printed from acrylonitrile butadiene styrene (ABS) plastic. The 3D printing fabrication process allowed for greater design freedom, and the final structure was both lightweight and rigid. A cross section view of the reflectometer is shown in Fig.\ref{reflectometer_xsec}. The illumination source for the reflectometer is a Horiba iHR-320 monochromator which is equipped with both a quartz tungsten halogen lamp (QTH) and a xenon arc lamp. The monochromator allows for the reflectance measurements to be performed through a range of wavelengths, and a slit to round Fiberguide fiber optic couples the monochromator to the reflectometer. The exit of the fiber illuminates a ground glass diffuser, which in turn illuminates a second ground glass diffuser. This pair of diffusers is intended to create a uniform illumination pattern, with the second diffuser being imaged by the pinhole. The pinhole is imaged by an aspheric lens, creating a collimated beam for the incident illumination of the sample. The required distance from the pinhole to the collimating lens is dependent upon wavelength. Over the intended operational bandpass of the reflectometer the aspheric lens to object distance varies by approximately 2 mm. To minimize the effects of the shifting focus distance aspheric lenses which are optimized for a narrower bandpasses were selected. The aspheric lenses were mounted at their required distances from the pinhole, and each has a broadband anti-reflective (AR) coating with bandpasses optimized for 350 nm - 700 nm, 650 nm - 1050 nm, and 1050 nm - 1700 nm. The collimated beam is reflected off the surface to be measured, and the reflection is collected by the signal photodiode. The active surface of the signal photodiode is 6.6 mm square and is overfilled by the 8 mm diameter collimated beam. The signal photodiode is mounted on an X-Y linear stage to allow for minor adjustments to the alignment of the photodiode to the collimated beam. The monitor diode is placed to measure the output of the second ground glass diffuser.

\begin{figure}
\centering
\begin{minipage}{.5\textwidth}
  \centering
  \includegraphics[width=.9\linewidth]{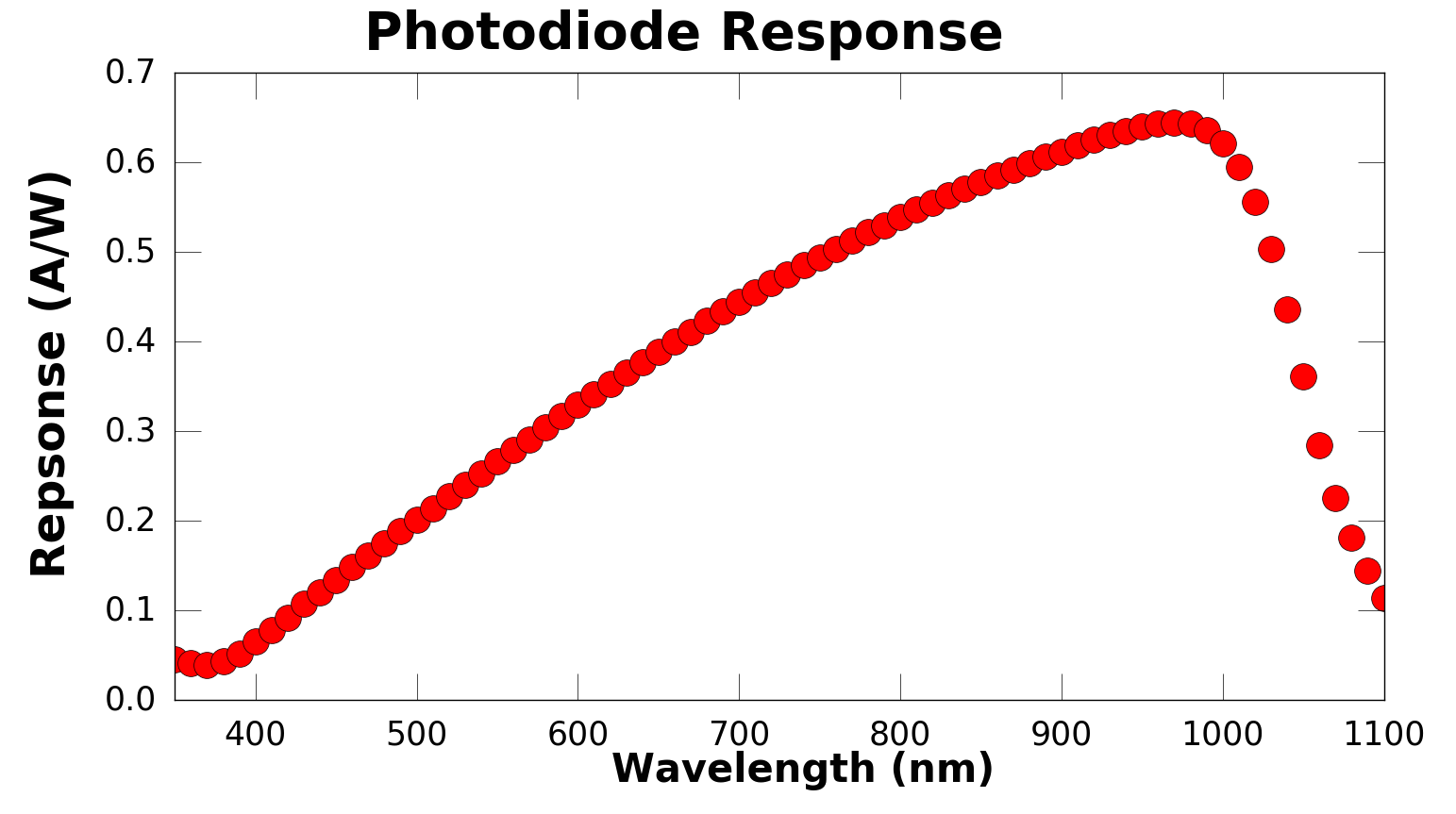}
  \captionsetup{width=.85\linewidth}
  \captionof{figure}{The response curve of the OSI PIN-44DPI photodiode.}
  \label{photodiode_response}
\end{minipage}%
\begin{minipage}{.5\textwidth}
  \centering
  \includegraphics[width=.9\linewidth]{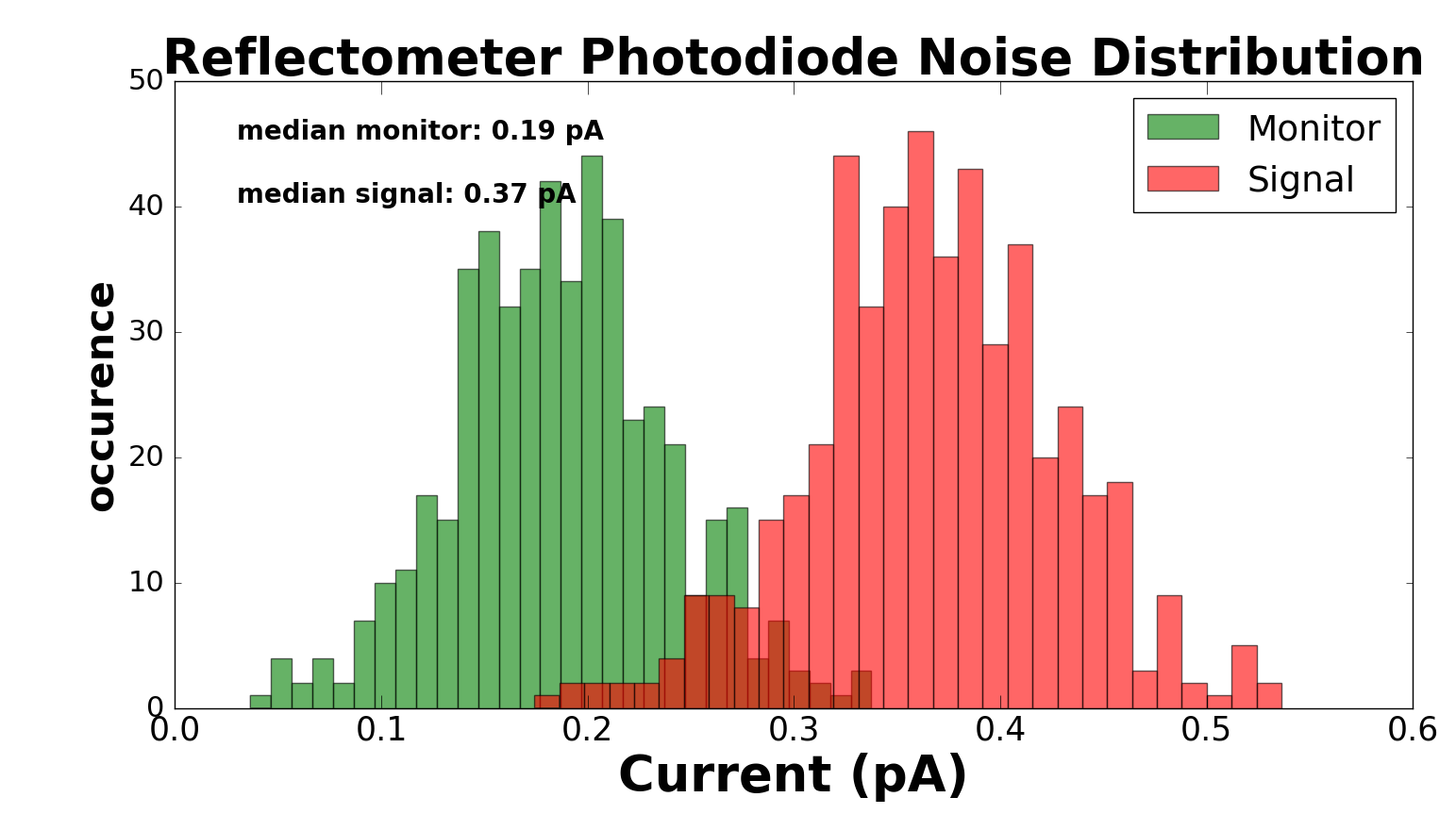}
  \captionsetup{width=.85\linewidth}
  \captionof{figure}{Noise distribution of the signal and monitor photodiodes.}
  \label{photodiode_noise}
\end{minipage}
\end{figure}

The photodiodes used for the signal and monitor measurements are OSI PIN-44DPI. The PIN-44DPI is a photovoltaic photodiode, which offers low noise at the expense of response speed. The response of the photodiode is shown in Fig.\ref{photodiode_response}. A Keithley 6482 dual channel pico-ammeter is used to measure the photodiode currents simultaneously. Fig.\ref{photodiode_noise} are the noise distributions for the entire signal chains of the signal and monitor photodiodes with zero illumination, and the noise does not significantly impact reflectance measurements. 

Reflectance measurements are performed in a Wenzel XO-87 coordinate measuring machine (CMM). A test cryostat is mounted in the CMM, and serves as the fixture for holding detectors. The reflectometer is attached to the CMM's ram, which allows for the reflectometer to be precisely positioned to perform measurements. Since the reflectance measurements are performed through the test cryostats window, and the reflectance of these devices can be as low as a few percent, the dewar windows can have a significant impact upon a reflectance measurement. The cryostat windows have been coated with a broadband AR coating to minimize the reflections from the window. 


\section{Reflectance Measurement Technique} 
\label{technique}

\begin{figure}[h]
	\hspace{1.2cm}
	\includegraphics[width=0.85\textwidth]{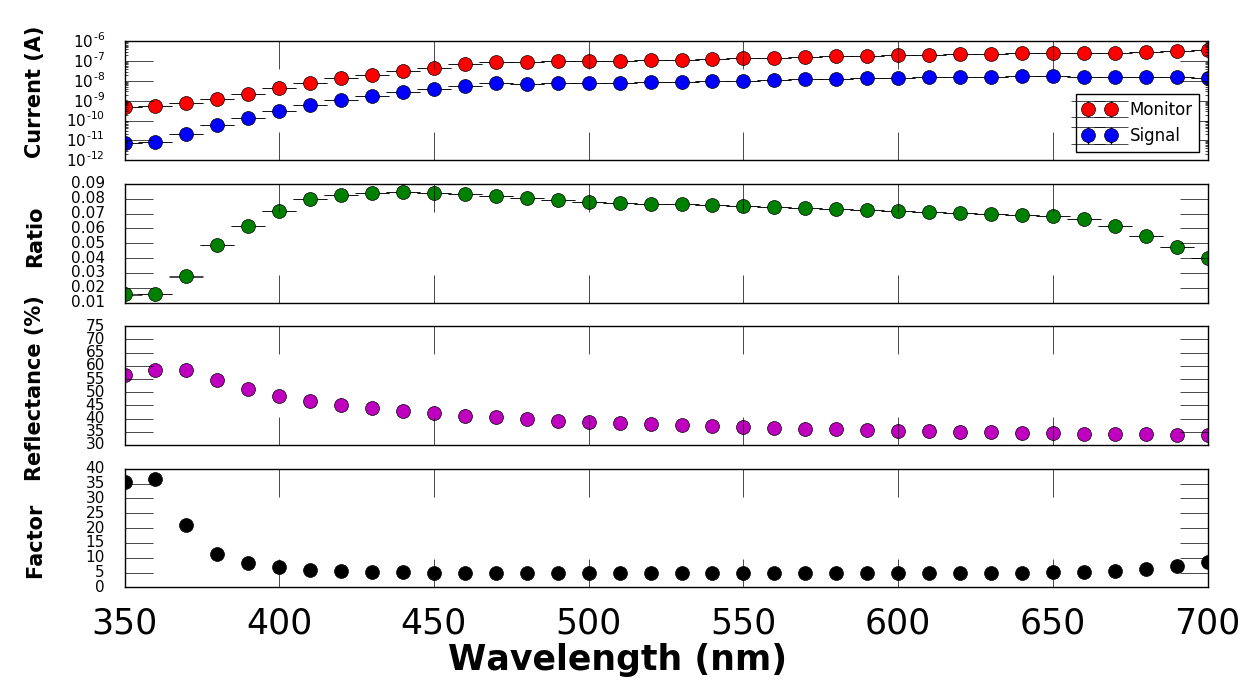}
	\caption{ Reflectance Measurement, the top panel shows the signal and monitor photodiode currents. The second panel is the ratio of the photodiode currents. The third panel is the calculated reflectance of silicon, and the bottom panel is the calibration factor.   }
	\label{calibration}
\end{figure}

Reflectance is the fraction of incident electromagnetic power that is not transmitted at an interface. The reflectometer uses a monitor diode to measure the incident power, and a signal diode to measure the reflected power. Using a monitor diode minimizes errors due to temporal variations in the illumination source. The top panel in Fig.\ref{calibration} are the measured monitor and signal currents for a reflectance measurement. A reflectance measurement is taken to be the ratio of the signal to monitor current, the second panel in Fig.\ref{calibration} is the ratio of the currents. For a typical reflectance measurement the incident power is approximately 100 times greater than the reflected power. 

The reflectometer measures reflectance relative to a silicon standard. The standard is a 2" silicon wafer which has been chemically mechanically polished (CMP) on one side. Since the reflectometer measures specular reflectance the calibration standard needs to have a similar surface roughness. Shen et al.\cite{Shen:2001tr} measured the bi-directional reflection distribution function (BRDF) for a range of silicon wafer roughnesses, and showed for surfaces where the Rayleigh criterion holds minimal amounts of light are diffusely scattered. By the Rayleigh criterion a surface is optically smooth if $\Delta h < \lambda / 8$, where $h$ is the RMS surface roughness and $\lambda$ is the wavelength of the incident light. While it is clear the CMP polished surface of the silicon standard is well below the Rayleigh criterion, the surface roughness of the device to be measured has not been quantified. But typically the wet chemical etching processes used to thin silicon decrease surface roughness, thus it is a reasonable assumption that the devices to be measured are also well below the Rayleigh criterion. 

Using the Fresnel equations, along with the indices of refraction ($\textit{n}$) and extinction coefficients ($\kappa$ ) for silicon taken from Green\cite{Green:2008dy}, the reflectance of silicon can be calculated. In Fig.\ref{calibration} the third panel shows the calculated reflectance of silicon. To derive a reflectance from a measured current ratio a wavelength dependent calibration factor is necessary. The calibration factor is found by dividing the silicon calculated reflectance by the silicon measured current ratio. The bottom panel in Fig.\ref{calibration} is an example of the calibration factor used to convert a current ratio into a reflectance. The reflectance of the device to be measured is found by multiplying the current ratio by the calibration factor. 

\begin{figure}[h]
	\hspace{4cm}
	\includegraphics[width=0.45\textwidth]{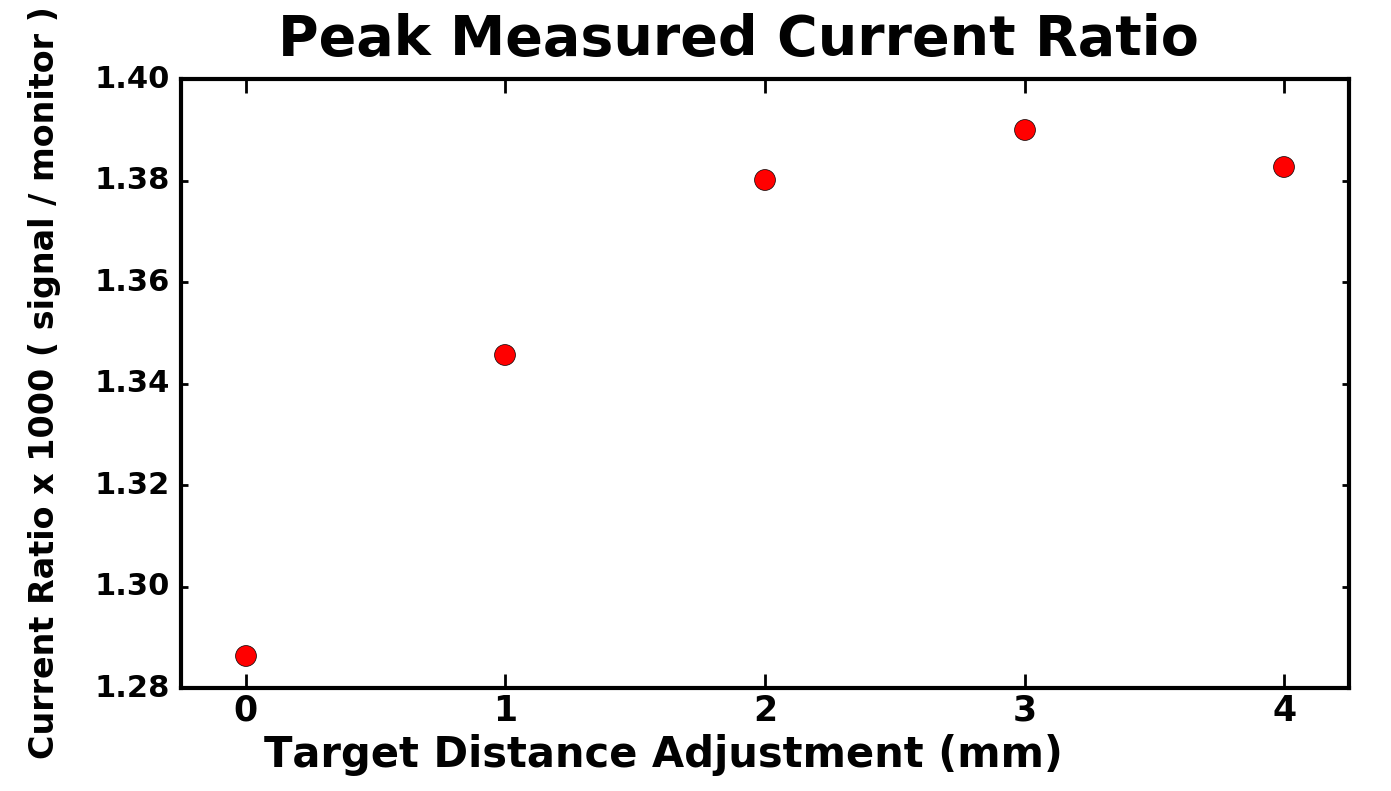}
	\caption{ Current ratio versus shift in distance to target. The peak signal is used to calculate the reflectance. }
	\label{target_distance}
\end{figure}

Relative reflectance measurements are susceptible to errors due to differing alignments of the calibration standard and the surfaces to be measured. Lesser\cite{Lesser:2014cx} uses an optical method by which to precisely align the surface to be tested to the same position as the calibration standard. In our reflectometer the distance to the target from the reflectometer affects where the collimated beam is reflected onto the signal photodiode. To overcome this limitation the CMM is used to measure reflectance through a range of target distances, with the peak signal taken to represent the measured reflectance. In Fig.\ref{target_distance} the current ratio is plotted versus shifts in distance to the target, and the peak signal is used to calculate the reflectance.

\section{CCD Reflectance Measurements} 
\label{results}

\begin{figure}
\centering
\begin{minipage}{.5\textwidth}
  \centering
  \includegraphics[width=.9\linewidth]{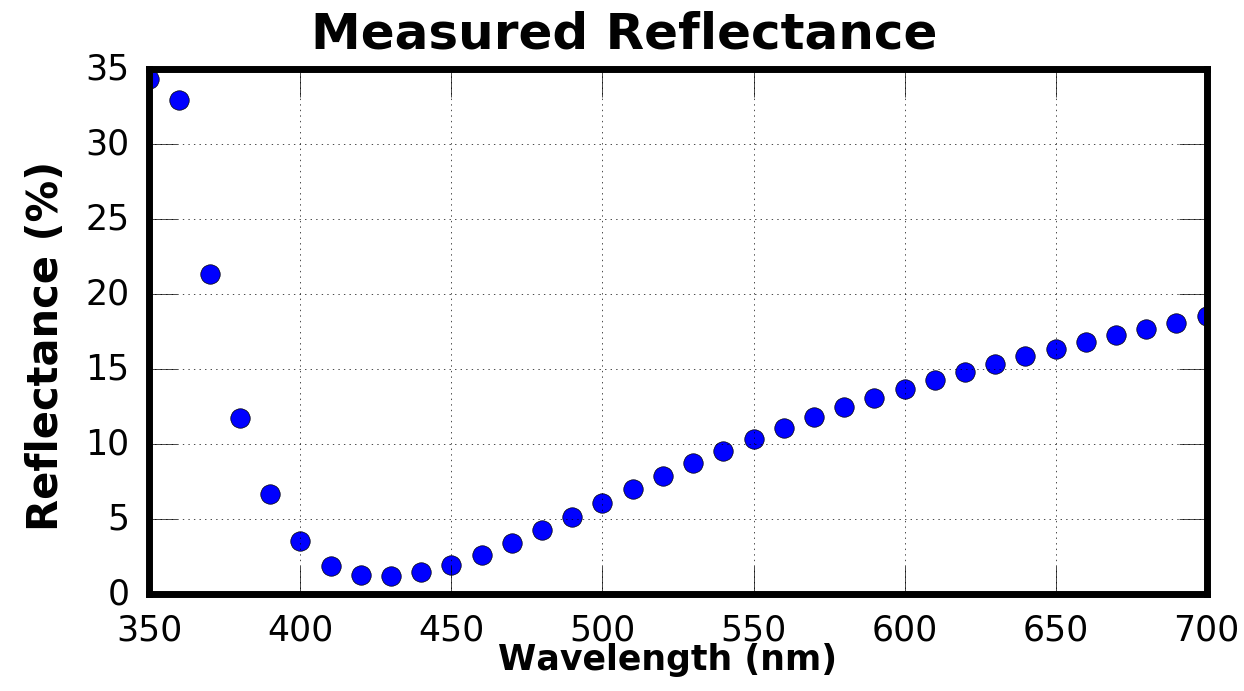}
  \captionsetup{width=.85\linewidth}
  \captionof{figure}{Measured reflectance of the blue optimized Hamamatsu CCD. }
  \label{B1_reflectance}
\end{minipage}%
\begin{minipage}{.5\textwidth}
  \centering
  \includegraphics[width=.9\linewidth]{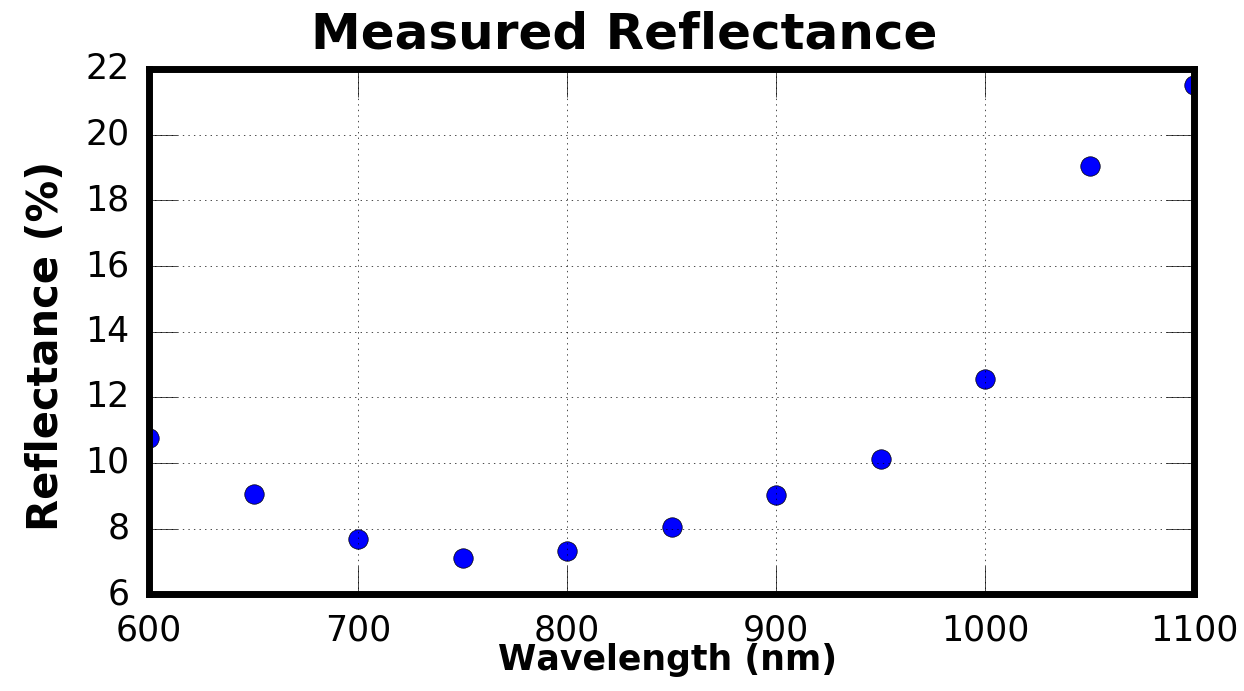}
  \captionsetup{width=.85\linewidth}
  \captionof{figure}{Measured reflectance of the red optimized Hamamatsu CCD. }
  \label{R1_reflectance}
\end{minipage}
\end{figure}

\begin{figure}
\centering
\begin{minipage}{.5\textwidth}
  \centering
  \includegraphics[width=.9\linewidth]{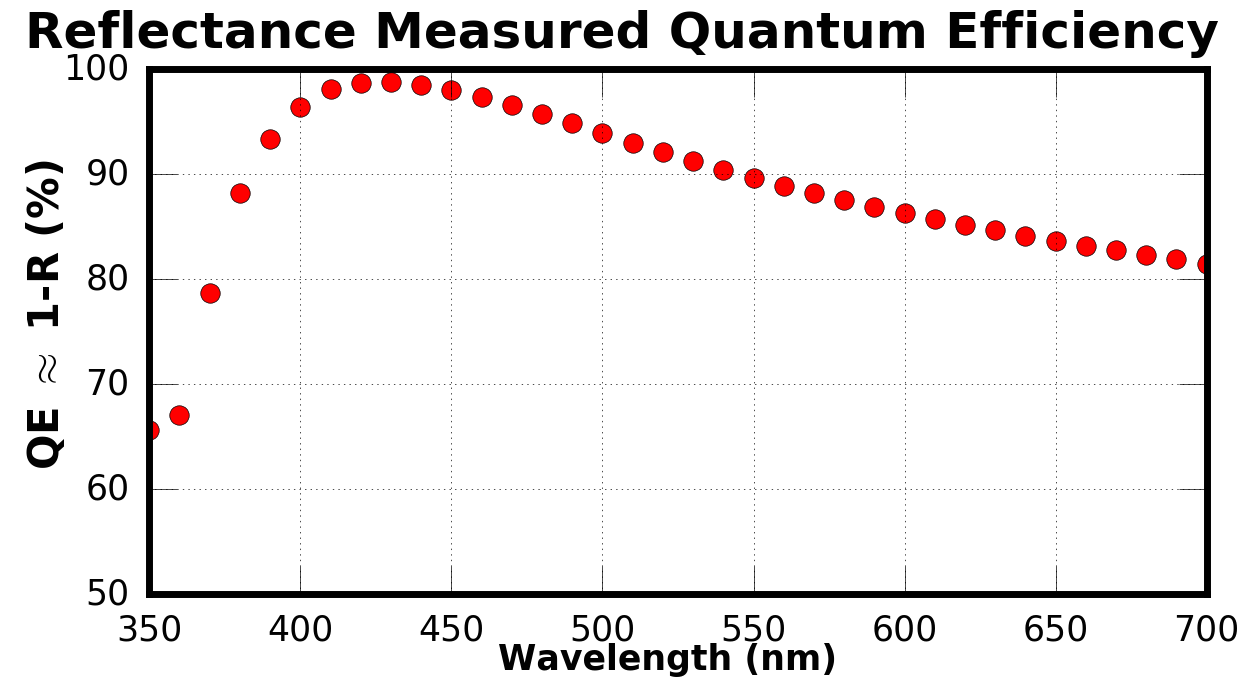}
  \captionsetup{width=.85\linewidth}
  \captionof{figure}{The reflectance measured quantum efficiency of the blue optimized Hamamatsu CCD. }
  \label{B1_QE}
\end{minipage}%
\begin{minipage}{.5\textwidth}
  \centering
  \includegraphics[width=.9\linewidth]{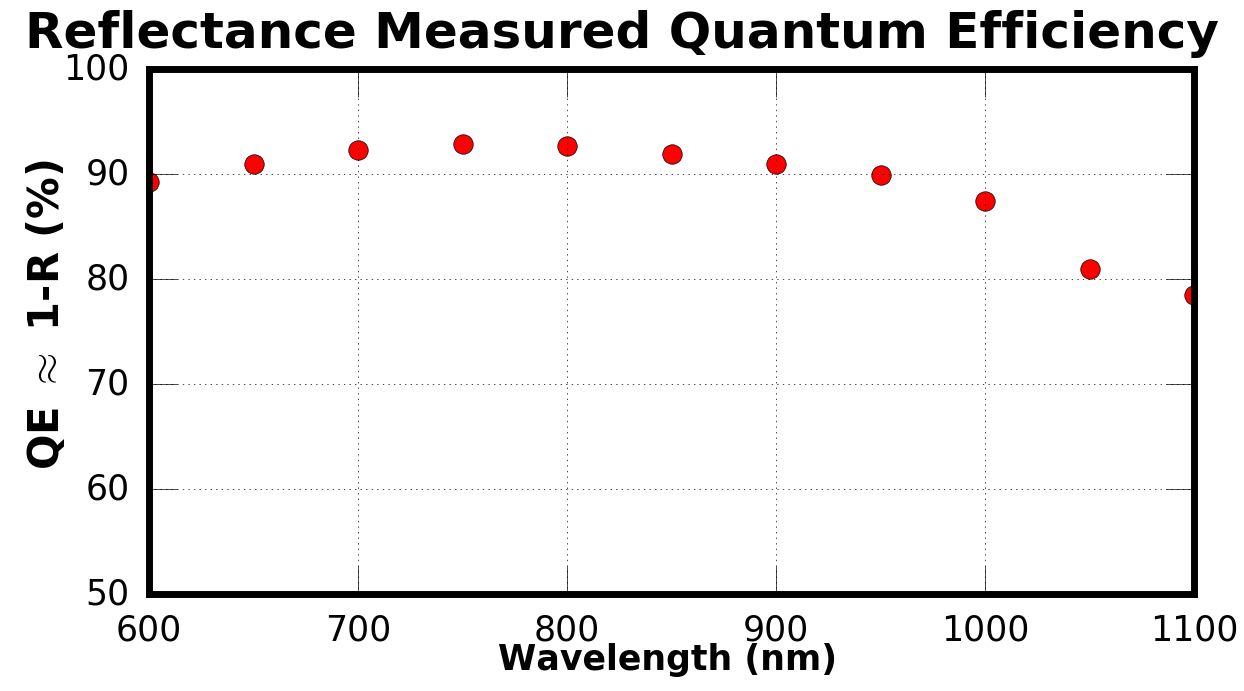}
  \captionsetup{width=.85\linewidth}
  \captionof{figure}{The reflectance measured quantum efficiency of the red optimized Hamamatsu CCD. }
  \label{R1_QE}
\end{minipage}
\end{figure}

Two types of Hamamatsu CCDs have been measured for their reflectance, a blue optimized CCD and red optimized CCD. The blue optimized CCD is a Hamamatsu part number S10892-1628(X), which is a backside illuminated fully depleted p-type device with a format of 4192 x 2048 15 $\mu$m pixels and is fabricated on 100-micron thick high-resistivity silicon with a broadband AR coating applied to the backside. These devices are detailed in Kamata et al.\cite{Kamata:2010ic} and Gunn et al.\cite{Gunn:2012wf}. The red optimized detector is a Hamamatsu S10892-1629(X), and is similar to the blue with the same geometry and mask, but it is fabricated on 200 $\mu$m thick high-resistivity silicon, and has an AR coating optimized for the red applied to the backside. Fig.\ref{B1_reflectance} is the measured reflectance of a blue optimized CCD, and Fig.\ref{R1_reflectance} is the measured reflectance of a red optimized CCD. Fabricius et al.\cite{Fabricius:2006uh} shows that between 500 - 900 nm the $QE=1-R$ and more generally $QE \le 1-R$. Fig.\ref{B1_QE} is the $1-R$ quantum efficiency for the blue optimized CCD, and Fig.\ref{R1_QE} is the $1-R$ quantum efficiency for the red optimized CCD. Kamata et al.\cite{Kamata:2010ic} measured the QE of the red optimized devices, and their QE curve is nearly identical to the reflectance measured QE from 600 nm - 950 nm. The intrinsic QE of the devices as measured by Kamata et al.\cite{Kamata:2010ic} begin to rapidly drop past 1,000 nm as the wavelengths approach the cutoff wavelength of silicon, and the measured reflectance is unable to capture this behavior. 

\begin{figure}[h!]
	\hspace{4cm}
	\includegraphics[width=0.65\textwidth]{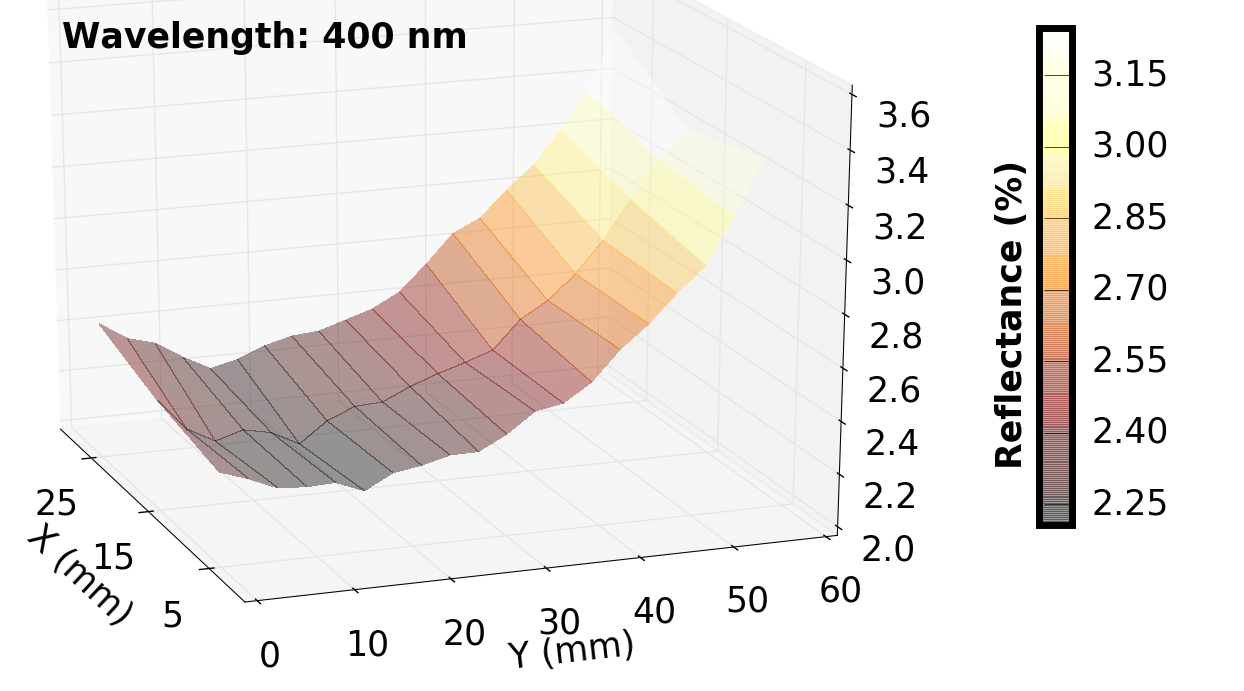}
	\caption{The surface reflectance of a blue optimized CCD taken at 400 nm. There is approximately a 1$\%$ reflectance gradient in the parallel direction.  }
	\label{B1_surface_reflectance}
\end{figure}

\begin{figure}[h!]
	\hspace{4cm}
	\includegraphics[width=0.65\textwidth]{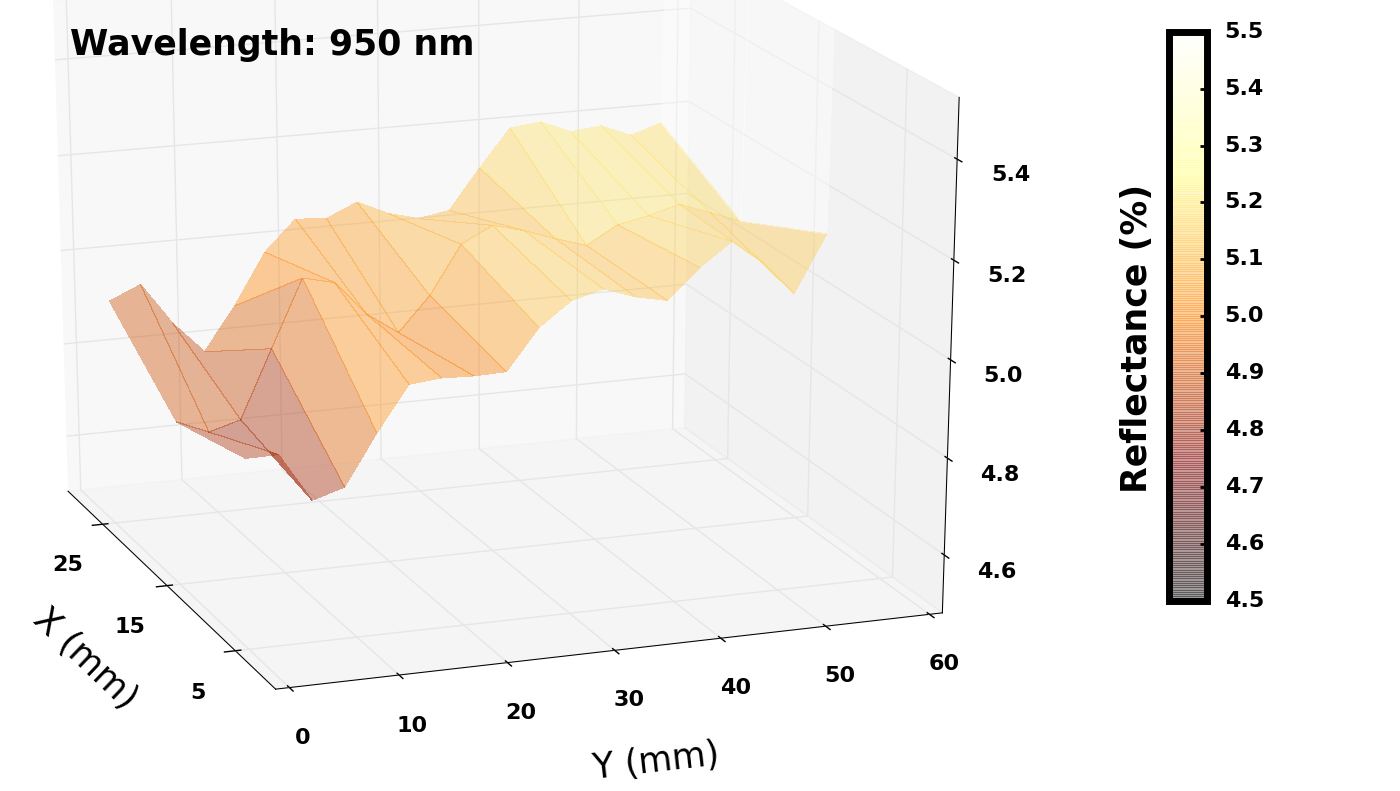}
	\caption{ The surface reflectance of a red optimized CCD taken at 950 nm. There is approximately a 0.5$\%$ reflectance gradient in the parallel direction. }
	\label{R1_surface_reflectance}
\end{figure}

The precise positioning ability of the CMM also allows for the measurement of the reflectance across the extent of a device to reveal reflectance gradients. The diameter of the reflectometer's collimated beam is 8 mm and the photodiode is 6.6 mm square, thus a reflectance measurement is sampling an area which is similar in extent to the photodiode. The Hamamatsu CCDs have an active area approximately 30 mm wide in the serial direction and 60 mm long in the parallel direction. The surface reflectance was sampled using three stripes spaced 10 mm apart, where a stripe runs in the parallel direction, and each stripe is sampled every 3 mm. The reflectance of the surfaces were measured through a range of wavelengths, with the blue optimized CCDs being measured from 350 nm - 700 nm at a 10 nm interval, and the red optimized CCDs were measured from 600 nm - 1,100 nm at a 50 nm interval. The reflectance was found to be a slowly varying function of wavelength. Fig.\ref{B1_surface_reflectance} is the reflectance across a blue optimized CCD taken at 400 nm. Fig.\ref{R1_surface_reflectance} is the reflectance across the surface of a red optimized CCD taken at 950 nm. Both CCDs show reflectance gradients in the parallel direction which has been attributed to the coating process, and the variation in reflectance in the serial direction is minimal. The shape and magnitude of the reflectance gradient is wavelength dependent.


\section{Conclusion} 
\label{conclusion}

We have built a novel reflectometer which used 3D printing to fabricate the super structure. The 3D printing process offers more design freedom, and dramatically reduces manufacturing times. By implementing a monitor and signal diode we have minimized measurement errors due to the temporal variations of the illumination source, and the relative reflectance method. The reflectometer measures reflectance relative to a silicon standard, and the errors due to the positioning differences between the standard and the device under test have been minimized by measuring the reflectance through a range of target distances and taking the peak value as the reflectance measurement. 

The reflectance of red and blue optimized Hamamatsu CCDs has been measured, and reflectance measured QE have been computed. The reflectance measured QE of the red optimized CCD matches the measured QE found for these devices given in other publications. We are currently preparing to measure the QE of the blue optimized devices, and we will then contrast the QE found by the two different techniques. 

Having the ability to measure the reflectance across the surface of the CCD has allowed us to quantify the reflectance gradient as a function of position and wavelength. We have measured a gradient in reflectance which is directed in the parallel direction of the CCDs, and the size and shape of the gradient has been found to be wavelength dependent. The reflectance gradient is attributed to the process by which the AR coating is applied.

\bibliography{reflectometer_bib}   

\begin{thebibliography}{1}

\bibitem{Shen:2001tr}
Y.~J. Shen, Z.~M. Zhang, B.~K. Tsai, and D.~P. Dewitt, ``{Bidirectional
  Reflectance Distribution Function of Rough Silicon Wafers},'' {\em
  International Journal of Thermophysics}~{\bf 22}, pp.~1311--1326, July 2001.

\bibitem{Green:2008dy}
M.~A. Green, ``{Self-consistent optical parameters of intrinsic silicon at 300K
  including temperature coefficients},'' {\em Solar Energy Materials and Solar
  Cells}~{\bf 92}, pp.~1305--1310, Nov. 2008.

\bibitem{Lesser:2014cx}
M.~Lesser, ``{Silicon sensor quantum efficiency, reflectance, and
  calibration},'' in {\em SPIE Astronomical Telescopes + Instrumentation},
  pp.~915411--1 -- 915411--12, SPIE, July 2014.

\bibitem{Kamata:2010ic}
Y.~Kamata, S.~Miyazaki, H.~Nakaya, H.~Suzuki, Y.~Miyazaki, and M.~Muramatsu,
  ``{Characterization and performance of hyper Suprime-Cam CCD},'' in {\em SPIE
  Astronomical Telescopes and Instrumentation: Observational Frontiers of
  Astronomy for the New Decade},  pp.~774229--774229--11, SPIE, 2010.

\bibitem{Gunn:2012wf}
J.~E. Gunn, M.~Carr, S.~A. Smee, J.~D. Orndorff, R.~H. Barkhouser, M.~Hart,
  C.~L. Bennett, J.~E. Greene, T.~Heckman, H.~Karoji, O.~LeFevre, H.-H. Ling,
  L.~Martin, B.~Menard, H.~Murayama, E.~Prieto, D.~Spergel, M.~A. Strauss,
  H.~Sugai, A.~Ueda, S.-Y. Wang, R.~Wyse, and N.~Zakamska, ``{Detectors and
  cryostat design for the SuMIRe Prime Focus Spectrograph (PFS)},'' {\em
  arXiv.org} , Oct. 2012.

\bibitem{Fabricius:2006uh}
M.~Fabricius, C.~J. Bebek, D.~E. Groom, A.~Karcher, and N.~Roe, ``{Quantum
  efficiency characterization of back-illuminated CCDs Part 2: reflectivity
  measurements},'' {\em SPIE}~{\bf 6068}, pp.~101--111, Jan. 2006.

\end{thebibliography}
\bibliographystyle{spiebib}

\end{document}